\documentclass[journal]{IEEEtran}
\IEEEoverridecommandlockouts
\usepackage{lineno}
\usepackage{hyperref}
\usepackage{cite}
\usepackage{amsmath,amssymb,amsfonts,cases} 
\usepackage{amsmath}
\usepackage{amsthm}
\DeclareMathOperator*{\argmax}{argmax}
\DeclareMathOperator*{\argmin}{argmin}
\usepackage{graphicx}
\usepackage{textcomp}
\usepackage{xcolor}
\usepackage{graphicx}
\usepackage{float}
\usepackage{subfigure}
\usepackage{amsmath,mleftright}
\usepackage{amsfonts,amssymb}
\usepackage{mathrsfs}
\usepackage{mathtools}
\usepackage{algorithm}
\usepackage{algorithmic}
\usepackage{bm}
\usepackage{multirow}
\usepackage{array}
\usepackage{amssymb}
\usepackage{amsmath}
\usepackage{cite}
\usepackage{url}
\usepackage{xcolor}
\usepackage{cite,graphicx,amsmath,amssymb}
\usepackage{subfigure}
\usepackage{fancyhdr}
\usepackage{mdwmath}
\usepackage{mdwtab}
\usepackage{caption}
\usepackage{amsthm}
\usepackage{setspace}
\usepackage{bm}
\usepackage{mathtools}
\usepackage{dsfont}
\usepackage{bbm}
\usepackage{framed}

\newtheorem{theorem}{Theorem}

\newtheorem{lemma}{Lemma}

\newtheorem{corollary}{Corollary}

\makeatletter
\newcommand{\biggg}{\bBigg@{3}}
\newcommand{\Biggg}{\bBigg@{3.5}}
\makeatother
\makeatletter
\renewcommand{\maketag@@@}[1]{\hbox{\m@th\normalsize\normalfont#1}}%
\makeatother
\def\BibTeX{{\rm B\kern-.05em{\sc i\kern-.025em b}\kern-.08em
    T\kern-.1667em\lower.7ex\hbox{E}\kern-.125emX}}
    \expandafter\def\expandafter\normalsize\expandafter{%
    \normalsize%
    \setlength\abovedisplayskip{4pt}%
    \setlength\belowdisplayskip{4pt}%
    \setlength\abovedisplayshortskip{2pt}%
    \setlength\belowdisplayshortskip{2pt}%
}
\allowdisplaybreaks[4]
\begin{document}
\title{Failure Detection for Pinching-Antenna Systems}
\author{Chongjun~Ouyang, Hao~Jiang, Zhaolin~Wang, Yuanwei~Liu, and Zhiguo~Ding\vspace{-10pt}
\thanks{C. Ouyang and H. Jiang are with the School of Electronic Engineering and Computer Science, Queen Mary University of London, London, E1 4NS, U.K. (e-mail: \{c.ouyang, hao.jiang\}@qmul.ac.uk).}
\thanks{Z. Wang and Y. Liu are with the Department of Electrical and Electronic Engineering, The University of Hong Kong, Hong Kong (email: \{zhaolin.wang, yuanwei\}@hku.hk).}
\thanks{Z. Ding is with the School of Electrical and Electronic Engineering (EEE), Nanyang Technological University, Singapore 639798 (zhiguo.ding@ntu.edu.sg).}}
\maketitle
\begin{abstract}
A signal processing-based framework is proposed for detecting random segment failures in segmented waveguide-enabled pinching-antenna systems. To decouple the passively combined uplink signal and to provide per-segment observability, tagged pilots are employed. A simple tag is attached to each segment and is used to apply a known low-rate modulation at the segment feed, which assigns a unique signature to each segment. Based on the tagged-pilot model, a low-complexity per-segment maximum-likelihood (ML) detector is developed for the case in which the pilot length is no smaller than the number of segments. For the case in which the pilot length is smaller than the number of segments, sparsity in the failure-indicator vector is exploited and a compressive sensing-based detector is adopted. Numerical results show that the per-segment detector approaches joint ML performance, while the compressive sensing-based detector achieves reliable detection with a short pilot and can outperform baselines that require much longer pilots.
\end{abstract}
\section{Introduction}
The pinching-antenna system (PASS) has attracted growing research interest as a reconfigurable-antenna architecture for next-generation networks \cite{liu2025pinching}. PASS, first proposed and prototyped by NTT DOCOMO \cite{suzuki2022pinching}, employs low-loss dielectric waveguides as signal conduits and deploys small dielectric particles, termed pinching antennas (PAs), along the waveguides. Each PA can radiate guided signals into free space and can couple free-space signals back into the waveguide \cite{liu2025pinching}. By adjusting the PA locations, PASS controls the phases and amplitudes of the radiated fields and achieves flexible beam-pattern shaping. PASS also supports waveguides with long spatial extent. This feature allows PAs to be placed close to users and enables strong line-of-sight (LoS) links, which alleviates large-scale path loss and mitigates blockage effects \cite{ding2024flexible,ding2025blockage}. In contrast, other reconfigurable-aperture technologies, such as movable antennas, fluid antennas, and reconfigurable intelligent surfaces, typically operate over apertures that span only a few to several tens of wavelengths \cite{ouyang2025array,samy2025pinching}. This extent is often insufficient at high-frequency bands envisioned for sixth-generation (6G) systems, such as the 7-24 GHz upper mid-bands \cite{bjornson2025enabling}.

Building on these advantages, recent works have optimized the placement of active PAs along a waveguide to enhance communication and sensing performance; see \cite{liu2025pinchingtutorial,liu2026survey} and references therein. Despite this progress, practical deployment remains challenging. The main benefit of PASS relies on long dielectric waveguides that sustain stable LoS links. A long monolithic waveguide, however, raises two key concerns. In-waveguide attenuation increases with waveguide length. Maintainability also becomes difficult since a failure can occur at an arbitrary location and repairs may require replacing a large portion of the structure, which increases maintenance cost and limits scalability.

\begin{figure}[!t]
\centering
    \subfigure[System setup.]
    {
        \includegraphics[width=0.45\textwidth]{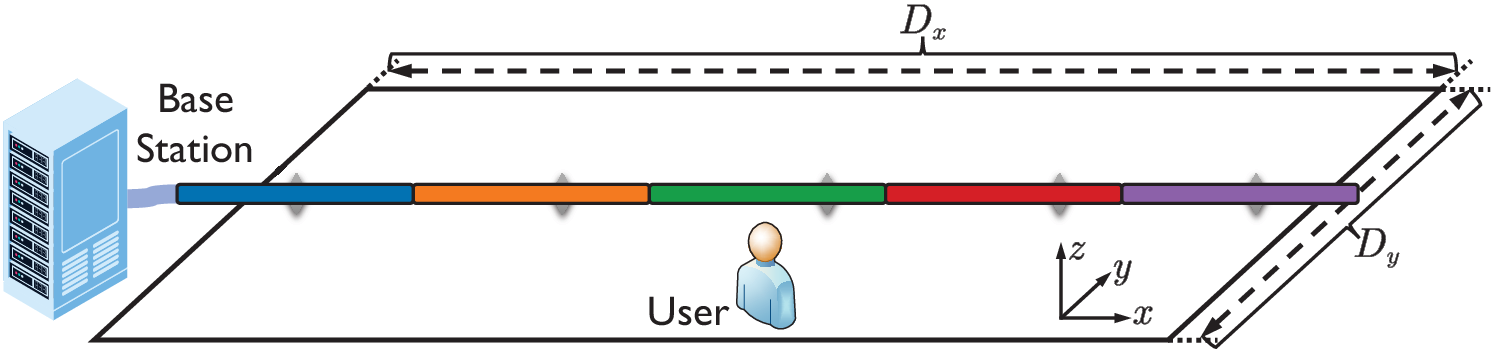}
	   \label{Figure: SWAN_System_Model}
    }
   \subfigure[Segmented waveguide.]
    {
        \includegraphics[width=0.45\textwidth]{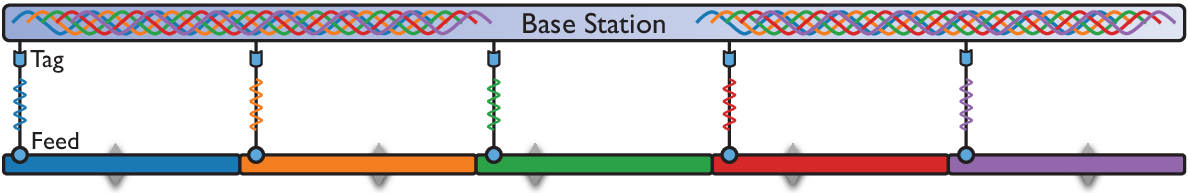}
	   \label{Figure: SWAN_System_Model1}
    }
\caption{Illustration of SWAN.}
\label{Figure: New_SWAN_System_Model}
\vspace{-10pt}
\end{figure}

To address these issues, a segmented waveguide-enabled pinching-antenna system (SWAN) was proposed in \cite{ouyang2025uplink}. SWAN employs multiple short dielectric waveguide segments that are arranged end-to-end without physical interconnection; see {\figurename} {\ref{Figure: New_SWAN_System_Model}}. Each segment has an independent feed point for signal injection and extraction. The extracted signals are forwarded to the base station (BS) through wired connections such as optical fiber or low-loss coaxial cables. Signal propagation and maintenance are confined within individual segments, which reduces in-waveguide attenuation and lowers repair cost.

In practice, each segment may experience random failures due to hardware defects, material degradation, or environmental factors. Reliable operation therefore requires timely identification of failed segments. A direct approach inspects segments sequentially, which is costly and time-consuming. This work developes a signal processing-based alternative for SWAN failure detection. We propose a tagged-pilot framework that attaches a simple tag to each segment and applies a known low-rate modulation at the segment feed, which provides a per-segment signature after passive combining at the BS. Based on this framework, we develop failure detection methods under two representative regimes. When the pilot length is no smaller than the number of segments, we derive a low-complexity per-segment maximum-likelihood (ML) detector that approaches the performance of joint ML detection. When the pilot length is smaller than the number of segments, we exploit sparse failures and design a detector based on compressive sensing. Numerical results validate the effectiveness of the proposed methods in terms of failure detection accuracy.

\section{System Model}\label{Section: System Model}
Consider a communication system where a BS employs a segmented pinched-waveguide structure to serve a single-antenna user, as shown in {\figurename} {\ref{Figure: SWAN_System_Model}}. The user is located in a rectangular service region with side lengths $D_x$ and $D_y$ along the $x$- and $y$-axes, respectively. The user location is denoted by ${\mathbf{u}}\triangleq[u_{x},u_{y},0]^{\mathsf{T}}$. The waveguide extends along the $x$-axis and covers the full horizontal span of the service region. It is partitioned into $M$ segments, each with length $L$, so that $D_x=LM$. 

Let ${\bm\psi}_{0}^{m}\triangleq[\psi_{0}^{m},0,d]^{\mathsf{T}}$ denote the feed-point location of the $m$th segment, where $\psi_{0}^{1}<\psi_{0}^{2}<\ldots<\psi_{0}^{M}$. The feed point is placed at the front-left end of each segment. One PA is deployed on each segment to radiate or capture signals \cite{suzuki2022pinching}. The PA location on the $m$th segment (i.e., the $m$th PA) is denoted by ${\bm\psi}^{m}\triangleq[\psi^{m},0,d]^{\mathsf{T}}$, and it satisfies
\begin{align}
\psi_{0}^{m}\leq \psi^{m}\leq \psi_{0}^{m}+L,\lvert\psi^{m}-\psi^{m'}\rvert\geq \Delta,\forall m\ne m',
\end{align}
where $\Delta>0$ is the minimum inter-antenna spacing used to mitigate mutual coupling \cite{ivrlavc2010toward}. 

The segment outputs are forwarded to the BS through wired links, such as optical fibers or low-loss coaxial cables, as shown in {\figurename} {\ref{Figure: New_SWAN_System_Model}}. These wired links are modeled as lossless because they incur negligible attenuation compared to the dielectric segments. Furthermore, prior studies indicate that in-waveguide propagation loss within each short segment remains negligible under typical SWAN parameters \cite{ouyang2025uplink}. Consequently, this impact is discarded in the subsequent analyses. All $M$ feed points connect to a single radio-frequency (RF) chain through a power combiner where the signals extracted from all segments are aggregated and forwarded for baseband processing.
\subsection{Signal Model}
Each waveguide segment may experience random failures due to hardware defects, material degradation, or environmental factors. Reliable transmission therefore requires timely identification of failed segments. To support failure detection, the user transmits a pilot sequence $x(t)$ with power $\lvert x(t)\rvert^2=P$ for $t=1,\ldots,T$. The signal received at the BS at time slot $t$ is expressed as follows:
\begin{equation}\label{Uplink_PASS_Basic_Model}
y(t)=\sum_{m=1}^{M}s_mh_{\rm{i}}({\bm\psi}_{0}^{m},{\bm\psi}^{m})
h_{\rm{o}}({\bm\psi}^{m},{\mathbf{u}})x(t)+n(t),
\end{equation}
where $h_{\rm{i}}({\bm\psi}_{0}^{m},{\bm\psi}^{m})$ and $h_{\rm{o}}({\bm\psi}^{m},{\mathbf{u}})$ denote the coefficients for the PA-to-feed in-waveguide propagation and the user-to-PA out-waveguide propagation, respectively, and $n(t)\sim{\mathcal{CN}}(0,\sigma^2)$ represents additive white Gaussian noise (AWGN) with variance $\sigma^2$. The binary random variable $s_m\in\{0,1\}$ characterizes the operational state of the $m$th segment for $m=1,\ldots,M$, where $s_m=1$ indicates a working segment and $s_m=0$ indicates a failed segment. 

Since PASS targets high-frequency bands \cite{suzuki2022pinching} where LoS propagation typically dominates \cite{ouyang2024primer}, we adopt the following free-space LoS model for the spatial channel coefficient between the PA and the user:
\begin{align}
h_{\rm{o}}({\bm\psi}^{m},{\mathbf{u}})\triangleq
\frac{\eta^{\frac{1}{2}}{\rm{e}}^{-{\rm{j}}k_0\lVert{\bm\psi}^{m}-{\mathbf{u}}\rVert}}{\lVert{\bm\psi}^{m}-{\mathbf{u}}\rVert},
\end{align}
where $\eta\triangleq\frac{c^2}{16\pi^2f_{\rm{c}}^2}$, $f_{\rm{c}}$ is the carrier frequency, $c$ is the speed of light, $k_0\triangleq\frac{2\pi}{\lambda_0}$ is the wavenumber, and $\lambda_0$ is the free-space wavelength. The in-waveguide propagation coefficient between the feed point and the PA is modeled as follows \cite{pozar2021microwave}:
\begin{align}\label{In_Waveguide_Channel_Model}
h_{\rm{i}}({\bm\psi}_{0}^{\rm{M}},{\bm\psi}^{\rm{M}})\triangleq{{\rm{e}}^{-\frac{2\pi}{\lambda_{\rm{g}}}\lVert{\bm\psi}_{0}^{\rm{M}}-{\bm\psi}^{\rm{M}}\rVert}}
={{\rm{e}}^{-\frac{2\pi}{\lambda_{\rm{g}}}\lvert{\psi}_{0}^{\rm{M}}-{\psi}^{\rm{M}}\rvert}},
\end{align}
where $\lambda_{\rm{g}}\triangleq\frac{\lambda}{n_{\rm{eff}}}$ is the guided wavelength and $n_{\rm{eff}}$ is the effective refractive index of the dielectric waveguide \cite{pozar2021microwave}.

Referring to \eqref{Uplink_PASS_Basic_Model}, the received samples can be expressed in the following vector form:
\begin{align}\label{Uplink_PASS_Basic_Model_Vector}
{\mathbf{y}}={\mathbf{x}}{\mathbf{h}}^{\mathsf{T}}{\mathbf{s}}+{\mathbf{n}},
\end{align}
where ${\mathbf{y}}\triangleq [y(1),\ldots,y(T)]^{\mathsf{T}}$ is the observation vector, ${\mathbf{s}}\triangleq [s_1,\ldots,s_M]^{\mathsf{T}}\in\{0,1\}^{M}\triangleq{\mathcal{S}}$ collects the operational states of all segments, ${\mathbf{x}}\triangleq[x(1),\ldots,x(T)]^{\mathsf{T}}$ is the pilot sequence, ${\mathbf{h}}\triangleq[h_1,\ldots,h_M]^{\mathsf{T}}$ is the effective channel vector with $h_m\triangleq h_{\rm{i}}({\bm\psi}_{0}^{m},{\bm\psi}^{m})h_{\rm{o}}({\bm\psi}^{m},{\mathbf{u}})$ for $m=1,\ldots,M$, and ${\mathbf{n}}\triangleq[n(1),\ldots,n(T)]^{\mathsf{T}}\sim{\mathcal{CN}}({\mathbf{0}},\sigma^2{\mathbf{I}}_T)$ is the noise vector. 

Assume that the channel information $\mathbf{h}$ is perfectly known at the BS. The objective is to infer ${\mathbf{s}}$ from ${\mathbf{y}}$. Direct recovery from \eqref{Uplink_PASS_Basic_Model_Vector} is, however, ill-posed. The effective measurement matrix for ${\mathbf{s}}$ is given by ${\mathbf{x}}{\mathbf{h}}^{\mathsf{T}}$, which has rank one. This rank deficiency prevents identification of ${\mathbf{s}}$ even in the absence of noise, since \eqref{Uplink_PASS_Basic_Model_Vector} reduces to a single scalar constraint ${\mathbf{h}}^{\mathsf{T}}{\mathbf{s}}=y(t)/x(t)$ for all $t$. As a result, multiple distinct state vectors $\mathbf{s}$ can yield the same observation ${\mathbf{y}}$. Physically, this ambiguity arises because the signals from all segments are aggregated passively into a single RF chain without distinct signatures. Consequently, it is fundamentally difficult to determine which segment fails from a single narrowband observation.
\subsection{Tagged-Pilot-Based Failure Detection}
To resolve the rank deficiency of the measurement function, we propose a tagging mechanism inspired by code-division multiple access (CDMA) principles. Specifically, we apply a unique low-rate modulation to the feed of each segment, thereby distinguishing their contributions within the combined RF chain. As illustrated in {\figurename} {\ref{Figure: SWAN_System_Model1}}, this is implemented by inserting a low-complexity control element, such as a PIN diode switch or a $0/\pi$ phase shifter, on the wired connection of each segment before the RF combiner. This hardware allows the $m$th segment's signal to be multiplied by a known sequence ${\mathbf{b}}_m\triangleq[b_m(1),\ldots,b_m(T)]^{\mathsf{T}}\in\{\pm1\}^{T}$ without altering the in-waveguide propagation. Consequently, the received signal at time slot $t$ is expressed as follows:
\begin{equation}\label{Uplink_PASS_Basic_Model}
y(t)=\sum\nolimits_{m=1}^{M}s_mb_m(t)h_mx(t)+n(t).
\end{equation}
Stacking the observations for $t = 1, \ldots, T$ yields the following vector form:
\begin{align}
{\mathbf{y}}={\mathsf{diag}}({\mathbf{x}}){\mathbf{B}}{\mathsf{diag}}({\mathbf{h}}){\mathbf{s}}+{\mathbf{n}},
\end{align}
where ${\mathbf{B}}\triangleq[{\mathbf{b}}_1,\ldots,{\mathbf{b}}_M]\in\{\pm1\}^{T\times M}$ is the tag matrix. In contrast to the untagged case, the effective measurement matrix ${\mathsf{diag}}({\mathbf{x}}){\mathbf{B}}{\mathsf{diag}}({\mathbf{h}})\in{\mathbbmss{C}}^{T\times M}$ now possesses a rank up to $\min\{T, M\}$, which provides the necessary degrees of freedom to recover $\mathbf{s}$. This structure allows all segments to be assessed simultaneously within a single pilot burst. For brevity, we assume a constant-modulus pilot $x(t) = \sqrt{P}$ for $t=1,\ldots,T$, which implies $\mathsf{diag}(\mathbf{x}) = \sqrt{P}\mathbf{I}_T$. It follows that
\begin{equation}\label{Tagged_Pilot_Model}
\mathbf{y} = \sqrt{P} \mathbf{B} \mathsf{diag}(\mathbf{h}) \mathbf{s} + \mathbf{n}.
\end{equation}

Conditioned on ${\mathbf{s}}$, the observation ${\mathbf{y}}$ follows a Gaussian distribution with mean $\sqrt{P}{\mathbf{B}}{\mathsf{diag}}({\mathbf{h}}){\mathbf{s}}$ and covariance $\sigma^2{\mathbf{I}}_T$. The likelihood function is given by
\begin{align}
p(\left.{\mathbf{y}}\right\rvert{\mathbf{s}})=\frac{1}{(\pi\sigma^2)^{T}}{\rm{e}}^{-\frac{1}{\sigma^2}\lVert{\mathbf{y}}-\sqrt{P}{\mathbf{B}}{\mathsf{diag}}({\mathbf{h}}){\mathbf{s}}\rVert^2}.
\end{align}
The ML estimate of ${\mathbf{s}}$ satisfies
\begin{align}\label{Joint_ML}
\hat{\mathbf{s}}=\argmax_{{\mathbf{s}}\in{\mathcal{S}}}p(\left.{\mathbf{y}}\right\rvert{\mathbf{s}})=
\argmin_{{\mathbf{s}}\in{\mathcal{S}}}\lVert{\mathbf{y}}-\sqrt{P}{\mathbf{B}}{\mathsf{diag}}({\mathbf{h}}){\mathbf{s}}\rVert^2.
\end{align}
This problem has exponential worst-case complexity in $M$. Tree-search methods such as sphere decoding can be applied, but the resulting complexity is prohibitive for large-scale segmented arrays. The recoverability also depends on the relationship between the number of segments $M$ and the pilot length $T$. Two cases are considered, namely overdetermined case $T \ge M$ and an underdetermined case $T < M$.
\section{Failure Detection for PASS}
\subsection{The Overdetermined Case}
Define ${\mathbf{a}}\triangleq {\mathsf{diag}}({\mathbf{h}}){\mathbf{s}}=[h_1s_1,\ldots,h_Ms_M]^{\mathsf{T}}$ and rewrite the tagged-pilot model in \eqref{Tagged_Pilot_Model} as follows:
\begin{align}
{\mathbf{y}}=\sqrt{P}{\mathbf{B}}{\mathbf{a}}+{\mathbf{n}}.
\end{align}
Since ${\mathbf{n}}\sim{\mathcal{CN}}({\mathbf{0}},{\mathbf{I}}_T)$, the ML estimate of $\mathbf{a}$ coincides with the following least-squares (LS) solution:
\begin{align}
\hat{\mathbf{a}}=
\argmin\nolimits_{\mathbf{a}}\lVert{\mathbf{y}}-\sqrt{P}{\mathbf{B}}{\mathbf{a}}\rVert^2.
\end{align} 
When $T\geq M$, the matrix ${\mathbf{B}}$ has full column rank, and thus the minimizer $\hat{\mathbf{a}}$ the minimizer is unique and satisfies the normal equation as follows:
\begin{align}\label{LS_General_Result}
{\mathbf{B}}^{\mathsf{H}}({\mathbf{y}}-\sqrt{P}{\mathbf{B}}\hat{\mathbf{a}})={\mathbf{0}}\Rightarrow
\hat{\mathbf{a}}=\frac{1}{\sqrt{P}}({\mathbf{B}}^{\mathsf{H}}{\mathbf{B}})^{-1}{\mathbf{B}}^{\mathsf{H}}{\mathbf{y}}.
\end{align}
The estimation error ${\mathbf{e}}\triangleq\hat{\mathbf{a}}-{\mathbf{a}}$ is complex Gaussian with mean zero and covariance ${\mathbf{C}}_{\mathbf{e}}\triangleq{\mathbbmss{E}}\{{\mathbf{e}}{\mathbf{e}}^{\mathsf{H}}\}=\frac{\sigma^2}{P}({\mathbf{B}}^{\mathsf{H}}{\mathbf{B}})^{-1}$. In general, ${\mathbf{C}}_{\mathbf{e}}$ is not diagonal when $\mathbf{B}$ is not orthogonal, so the estimation errors across segments are correlated. A low-complexity detector decides each segment state from the scalar hypotheses ${a}_m\in\{0,h_m\}$. Let $\sigma_m^2$ denote the $m$th diagonal entry of ${\mathbf{C}}_{\mathbf{e}}$. The per-segment hypotheses are given by
\begin{align}
\left\{\begin{array}{l}
{\mathcal{H}}_0:s_m=0\Rightarrow\hat{a}_m\sim{\mathcal{CN}}(0,\sigma_m^2),\\
{\mathcal{H}}_1:s_m=1\Rightarrow\hat{a}_m\sim{\mathcal{CN}}(h_m,\sigma_m^2).
\end{array}\right.
\end{align}
Since both hypotheses share the same variance, the per-segment ML rule can be written as follows:
\begin{equation}\label{Per_Segment}
\Re\{h_m^{*}\hat{a}_m\}
\underset{H_0}{\overset{H_1}{\gtrless}} \frac{\lvert h_m\rvert^2}{2},
\end{equation}
where ${\mathbf{a}}=[\hat{a}_1,\ldots,\hat{a}_M]^{\mathsf{T}}$, and it avoids the tree search in \eqref{Joint_ML}. 

The tag matrix ${\mathbf{B}}$ affects the estimation accuracy through ${\mathbf{B}}^{\mathsf{H}}{\mathbf{B}}$. Under the equal-energy constraint $\lVert{\mathbf{b}}_m\rVert^2=T$ for all columns, a standard objective is to minimize the total LS error ${\mathbbmss{E}}\{\lVert \hat{\mathbf{a}}-{\mathbf{a}}\rVert^2\}={\mathsf{tr}}({\mathbf{C}}_{\mathbf{e}})=\frac{\sigma^2}{P}{\mathsf{tr}}(({\mathbf{B}}^{\mathsf{H}}{\mathbf{B}})^{-1})$. This objective is minimized when the columns of ${\mathbf{B}}$ are mutually orthogonal, which yields ${\mathbf{B}}^{\mathsf{H}}{\mathbf{B}}={{T}}{\mathbf{I}}_M$. It follows that $\hat{\mathbf{a}}=\frac{1}{T\sqrt{P}}{\mathbf{B}}^{\mathsf{H}}{\mathbf{y}}$, and inter-segment interference is removed. In this case, the per-segment decision rule in \eqref{Per_Segment} matches the joint ML detector in \eqref{Joint_ML}. A convenient $\{\pm1\}$-valued orthogonal construction uses Walsh-Hadamard sequences. Let $T=2^{\lceil\log_2{M}\rceil}$ and let ${\mathbf{H}}_T\in\{\pm1\}^{T\times T}$ denote the Walsh-Hadamard matrix. The tag matrix can be constructed by selecting $M$ rows from ${\mathbf{H}}_T$, which satisfies ${\mathbf{B}}^{\mathsf{H}}{\mathbf{B}}={{T}}{\mathbf{I}}_M$. For an arbitrary pilot length $T$ that does not admit a square Walsh-Hadamard matrix, a practical rule is to reduce inter-column correlation. This can be achieved by selecting a $T\times M$ submatrix from a larger Walsh-Hadamard matrix ${\mathbf{H}}_{T_0}$ with $T_0=2^{\lceil\log_2{\max\{M,T\}}\rceil}$. A random selection yields near-orthogonal columns in practice.
\subsection{The Underdetermined Case}
A wide service region often requires a large number of waveguide segments, which leads to an underdetermined regime with $M>T$, and may yield $M\gg T$ in dense deployments. In this case, the LS estimator in \eqref{LS_General_Result} is not applicable since ${\mathbf{B}}$ does not have full column rank and $({\mathbf{B}}^{\mathsf{H}}{\mathbf{B}})^{-1}$ does not exist. Direct recovery of the binary state vector $\mathbf{s}$ from ${\mathbf{y}}$ is therefore ill-posed.

A practical property is that only a small fraction of segments change state over a short monitoring interval. Since failures across segments are typically independent, the probability of a large number of simultaneous failures is low. This motivates a \emph{sparse}-change formulation. Let ${\mathbf{s}}_0\in{\mathcal{S}}$ denote a reference operational-state vector that represents the most recent verified state of the segmented waveguide, e.g., obtained during commissioning or from the previous monitoring instant when no alarm was raised. Define the failure-indicator vector ${\mathbf{f}}\triangleq[f_1,\ldots,f_M]^{\mathsf{T}}={\mathbf{s}}_0-{\mathbf{s}}\in\{0,1\}^{M}$, where $f_m=1$ indicates that the $m$th segment has failed relative to ${\mathbf{s}}_0$. 

Substituting ${\mathbf{s}}={\mathbf{s}}_0-{\mathbf{f}}$ into \eqref{Tagged_Pilot_Model} gives
\begin{align}
{\mathbf{y}}=-{\mathbf{A}}_{\mathbf{f}}{\mathbf{s}}_0+{\mathbf{A}}_{\mathbf{f}}{\mathbf{f}}+{\mathbf{n}},
\end{align}
where ${\mathbf{A}}_{\mathbf{f}}\triangleq-{\mathsf{diag}}({\mathbf{x}}){\mathbf{B}}{\mathsf{diag}}({\mathbf{h}})$. The BS removes the known reference contribution and forms the residual as follows:
\begin{align}\label{LASSO_Model_Pre}
{\mathbf{r}}\triangleq{\mathbf{y}}+{\mathbf{A}}_{\mathbf{f}}{\mathbf{s}}_0={\mathbf{A}}_{\mathbf{f}}{\mathbf{f}}+{\mathbf{n}}.
\end{align}
The unknown ${\mathbf{f}}$ is real-valued and nonnegative, and it is \emph{sparse} when only a few segments change state between two monitoring instants. Since $T<M$, recovering $\mathbf{f}$ from $\{{\mathbf{r}},{\mathbf{A}}\}$ admits infinitely many solutions unless additional structure is imposed. We therefore adopt an $\ell_1$-regularized estimator based on the \emph{least absolute shrinkage and selection operator (LASSO)}, which promotes sparsity while fitting the measurements. To apply standard solvers, we convert the complex model into an equivalent real-valued system while keeping a real unknown. Define $\tilde{{\mathbf{r}}}\triangleq[\Re\{{\mathbf{r}}^{\mathsf{T}}\},\Im\{{\mathbf{r}}^{\mathsf{T}}\}]^{\mathsf{T}}$ and $\tilde{\mathbf{A}}\triangleq[\Re\{{\mathbf{A}}^{\mathsf{T}}\},\Im\{{\mathbf{A}}^{\mathsf{T}}\}]^{\mathsf{T}}$, which transformers \eqref{LASSO_Model_Pre} into the following form:
\begin{align}
\tilde{{\mathbf{r}}}=\tilde{\mathbf{A}}{\mathbf{f}}+\tilde{{\mathbf{n}}},
\end{align}
where $\tilde{{\mathbf{n}}}\triangleq[\Re\{{\mathbf{n}}^{\mathsf{T}}\},\Im\{{\mathbf{n}}^{\mathsf{T}}\}]^{\mathsf{T}}\sim{\mathcal{N}}({\mathbf{0}},\frac{\sigma^2}{2}{\mathbf{I}}_{2T})$. This transformation uses both the in-phase and quadrature observations, but it does not introduce an imaginary part for the unknown vector. The failure indicator ${\mathbf{f}}$ is then estimated by solving the following convex problem: 
\begin{align}
\hat{\mathbf{f}}=\argmin\nolimits_{{\mathbf{f}}}\frac{1}{2}\lVert\tilde{{\mathbf{r}}}-\tilde{\mathbf{A}}{\mathbf{f}}\rVert^2+\lambda\lVert{\mathbf{f}}\rVert_1,
\end{align}
where $\lambda>0$ controls the sparsity level. This problem can be implemented efficiently using MATLAB's built-in \verb"lasso" function applied to $(\tilde{\mathbf{A}},\tilde{{\mathbf{r}}})$. In our implementation, we compute a solution path over a set of $\lambda$ values and select $\lambda$ using a discrepancy-type rule, i.e., the $\lambda$ for which the residual energy $\lVert\tilde{{\mathbf{r}}}-\tilde{\mathbf{A}}\hat{\mathbf{f}}\rVert^2$ is closest to the expected noise energy $T\sigma^2$. After obtaining $\hat{\mathbf{f}}=[\hat{f}_1,\ldots,\hat{f}_M]^{\mathsf{T}}$, the segment states are declared by a scalar threshold test. Since ${{f}}_m\in\{0,1\}$, we decide each $\hat{f}_m$ for $m=1,\ldots,M$ as follows:
\begin{equation}
\hat{f}_m
\underset{f_m=0}{\overset{f_m=1}{\gtrless}} \tau,
\end{equation}
where $\tau\in(0,1)$ is a threshold and $\tau=0.5$ is a natural choice under the binary model. This procedure enables simultaneous monitoring of all segments with a short pilot burst even when $M\gg T$, and it avoids exhaustive search over ${\mathbf{s}}$.
\section{Numerical Results}
Numerical results are presented to validate the proposed failure detection methods. Unless stated otherwise, we set $f_{\rm{c}} = 28$ GHz, $n_{\rm{eff}} = 1.4$, $\Delta = \frac{\lambda_0}{2}$, and $\sigma^2=-90$ dB. The user is located at the origin, so $u_x=u_y=0$ m. The waveguide is deployed at height $d = 3$ m. The first segment feed point is placed at $\psi_{0}^{1}=-\frac{D_x}{2}$. Each segment has length $L=1$ m. The failure probability of each segment is set to $0.02$. All results are averaged over $10^5$ independent realizations of the operational-state vector ${\mathbf{s}}$.

\begin{figure}[!t]
\centering
\includegraphics[width=0.45\textwidth]{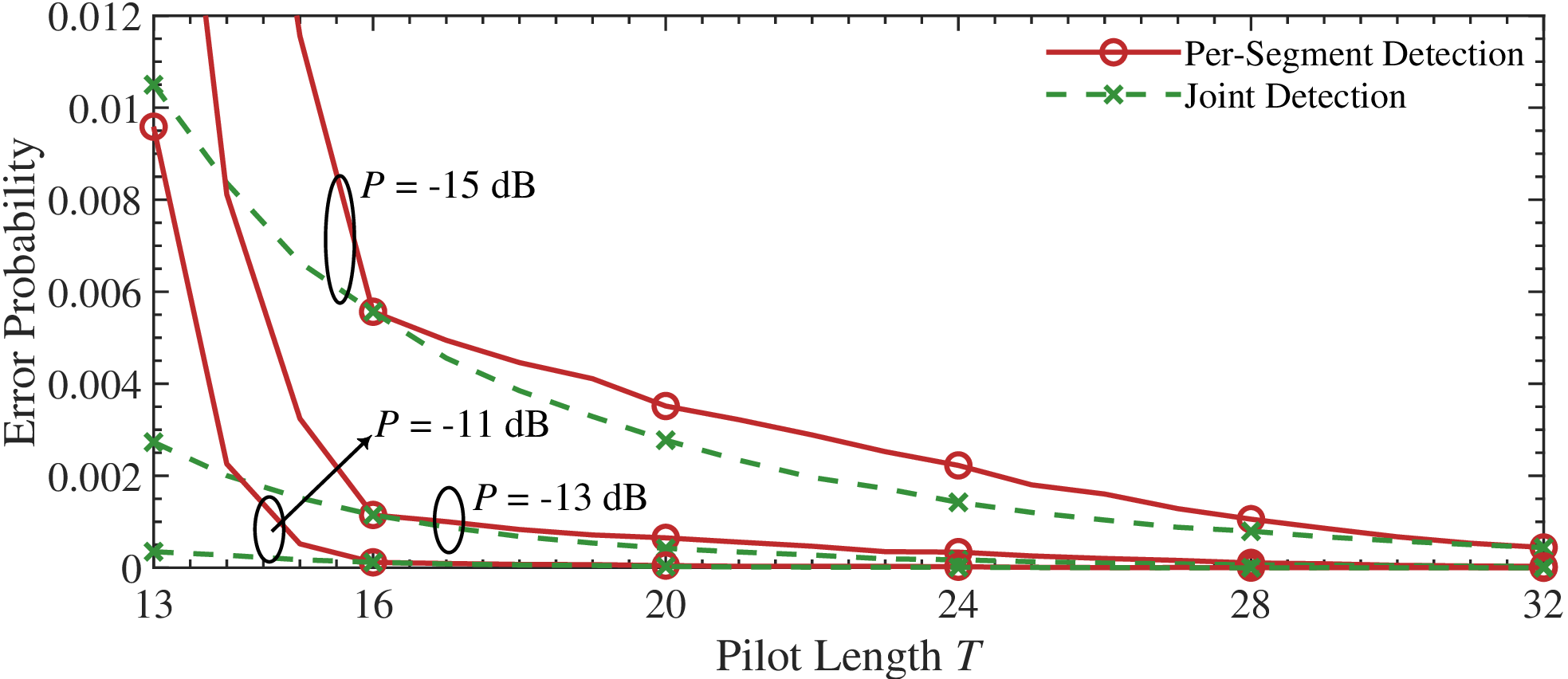}
\caption{The overdetermined case. $M=13$.}
\label{Figure_Failure_Detection_Overdetermined_Fig2}
\vspace{-10pt}
\end{figure}

\begin{figure}[!t]
\centering
\includegraphics[width=0.45\textwidth]{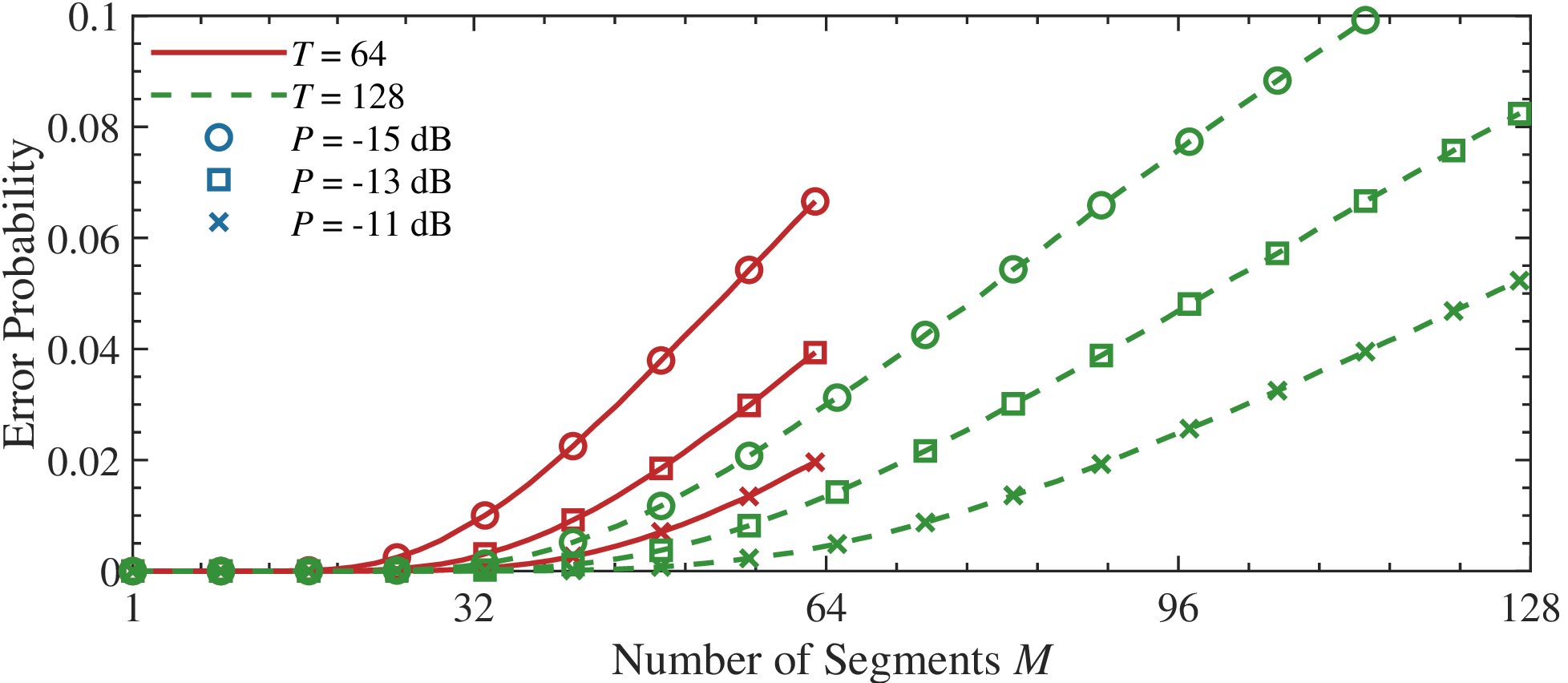}
\caption{The overdetermined case.}
\label{Figure_Failure_Detection_Overdetermined_Fig1}
\vspace{-10pt}
\end{figure}

{\figurename} {\ref{Figure_Failure_Detection_Overdetermined_Fig2}} shows the failure detection error probability versus the pilot length under the overdetermined case for selected transmit powers. The per-segment ML detector in \eqref{Per_Segment} is compared with the joint ML detector in \eqref{Joint_ML}. As expected, the error probability decreases as the transmit power increases, and joint ML detection provides the best performance. The performance gap between joint detection and per-segment detection is observed to shrink as the pilot length increases, since a longer pilot provides more degrees of freedom for the tagged matrix and reduces inter-segment correlation. When the pilot length equals a power of two, the tag matrix can be constructed from Walsh-Hadamard sequences, which yields an orthogonal design. In this case, the per-segment detector matches the joint ML detector, as also observed at $T=16$.

\begin{figure}[!t]
\centering
\includegraphics[width=0.45\textwidth]{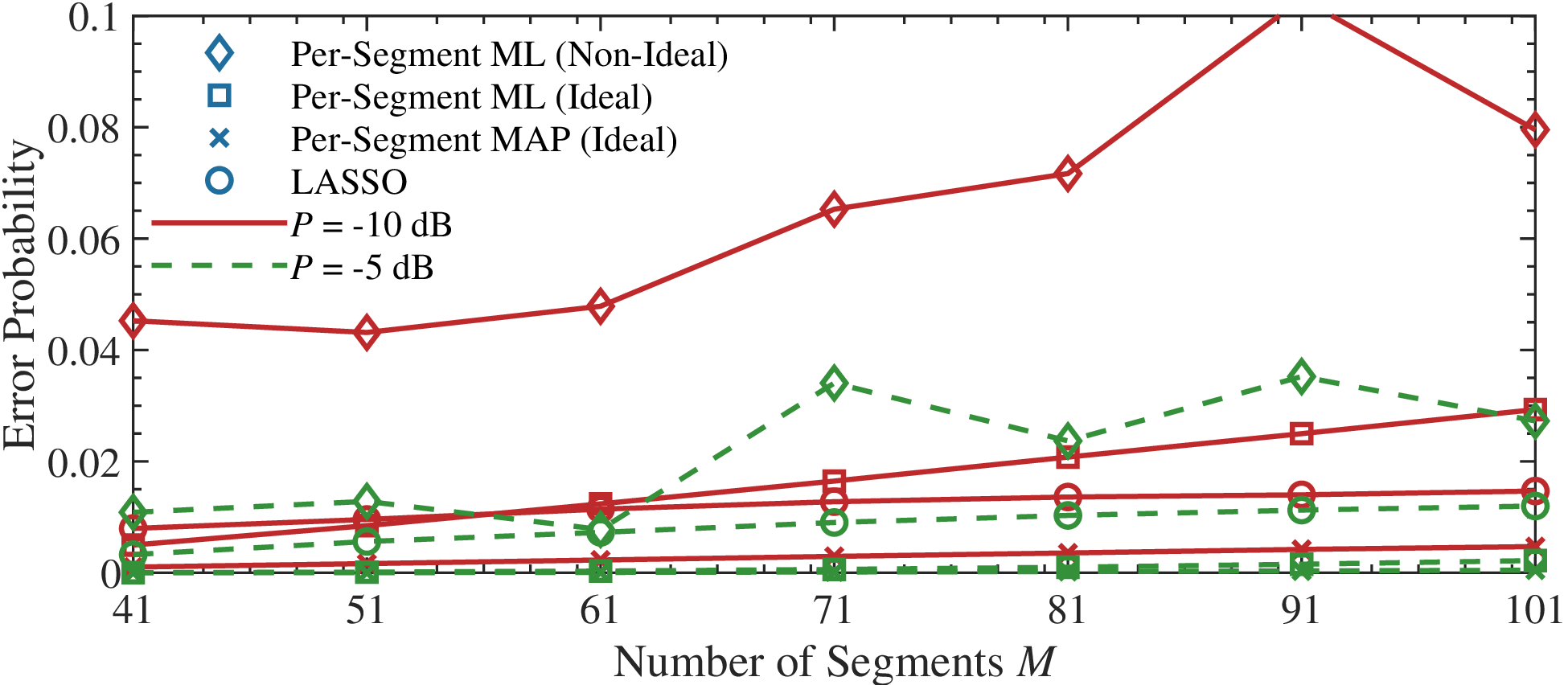}
\caption{The underdetermined case. $T=32$ and ${\mathbf{s}}_0={\mathbf{1}}$.}
\label{Figure_Failure_Detection_Underdetermined_LASSO_Fig1}
\vspace{-10pt}
\end{figure}

{\figurename} {\ref{Figure_Failure_Detection_Overdetermined_Fig1}} shows the error probability versus the number of segments under the overdetermined case for selected pilot lengths. The pilot lengths are chosen as powers of two, so the per-segment detector achieves the same performance as joint ML detection. It is observed that the error probability increases with the number of segments when the pilot length is fixed. This trend reflects the increased difficulty of distinguishing failures across a larger array under a limited pilot budget. Furthermore, it can be seen from this graph that a longer pilot reduces the error probability by improving the conditioning of the tagged measurement matrix.

{\figurename} {\ref{Figure_Failure_Detection_Underdetermined_LASSO_Fig1}} presents the error probability in the underdetermined case with a short pilot. The proposed LASSO-based detector is compared with two probing baselines that use a pilot length equal to the number of segments, i.e., $T=M$. In the first baseline, the tag matrix is formed by selecting $T$ rows from a larger Walsh-Hadamard matrix, which yields a non-orthogonal design. In the second baseline, an ideal orthogonal design is considered by setting ${\mathbf{B}}=\sqrt{T}{\mathbf{I}}_T$. As expected, the LASSO detector benefits from higher transmit power. Despite using a much shorter pilot, the LASSO detector consistently outperforms the probing baseline with non-orthogonal tags, and it can outperform the ideal linear probing baseline in certain transmit power regimes, since sparsity in the failure-indicator vector is exploited while noise and inter-segment interference are suppressed. An oracle probing scheme that combines ${\mathbf{B}}=\sqrt{T}{\mathbf{I}}_T$ with a \emph{maximum a posteriori (MAP)} detector using the prior failure probability $0.02$ provides a lower bound. The LASSO detector remains a low-overhead solution that can closely approach this bound in the underdetermined case.
\section{Conclusion}
This article has proposed a signal-processing-based failure detection framework for SWAN using tagged pilots. In the overdetermined case, the derived per-segment ML detector approaches the performance of joint ML detection while maintaining low complexity. In the underdetermined case, the proposed LASSO-based detector exploits sparse failures and achieves reliable detection with a short pilot. Numerical results verify the effectiveness of the proposed methods and indicate that tagged-pilot-based monitoring provides a useful baseline for developing more advanced failure diagnosis and maintenance strategies in SWAN deployments.
\clearpage
\bibliographystyle{IEEEtran}
\bibliography{mybib}
\end{document}